\documentclass[letterpaper,twocolumn,10pt]{article}
\usepackage{usenix-2020-09}

\usepackage{tikz}
\usepackage{amsmath}
\usepackage{algorithm}
\usepackage{algorithmic}
\usepackage{booktabs}
\usepackage{multirow}
\usepackage{xcolor}
\usepackage{url}
\usepackage{enumitem}

\makeatletter
\def\@fnsymbol#1{\ensuremath{\ifcase#1\or \textcolor{lightgray}{*}\or 
\textcolor{lightgray}{\dagger}\or \textcolor{lightgray}{\ddagger}\or 
\textcolor{lightgray}{\mathsection}\or \textcolor{lightgray}{\mathparagraph}\or 
\textcolor{lightgray}{**}\or \textcolor{lightgray}{\dagger\dagger}\or 
\textcolor{lightgray}{\ddagger\ddagger}\else\@ctrerr\fi}}
\makeatother

\usepackage{tikz}
\usetikzlibrary{positioning,arrows.meta,fit,calc}

\begin{document}

\date{}

\title{\Large \bf Multi-Agent Penetration Testing AI for the Web}

\author{
{\rm Isaac David}\\
University College London
\and
{\rm Arthur Gervais}\\
University College London
}

\maketitle

\begin{abstract}
AI-powered development platforms are making software creation accessible to a broader audience, but this democratization has triggered a scalability crisis in security auditing. With studies showing that up to 40\% of AI-generated code contains vulnerabilities~\cite{pearce2022asleep}, the pace of development now vastly outstrips the capacity for thorough security assessment.

We present \textsc{MAPTA}, a multi-agent system for autonomous web application security assessment that combines large language model orchestration with tool-grounded execution and end-to-end exploit validation. On the 104-challenge XBOW benchmark, \textsc{MAPTA} achieves 76.9\% overall success with perfect performance on SSRF and misconfiguration vulnerabilities, 83\% success on broken authorization, and strong results on injection attacks including server-side template injection (85\%) and SQL injection (83\%). Cross-site scripting (57\%) and blind SQL injection (0\%) remain challenging. Our comprehensive cost analysis across all challenges totals \$21.38 with a median cost of \$0.073 for successful attempts versus \$0.357 for failures. Success correlates strongly with resource efficiency, enabling practical early-stopping thresholds at approximately 40 tool calls or \$0.30 per challenge.

MAPTA's real-world findings are impactful given both the popularity of the respective scanned GitHub repositories (8K-70K stars) and MAPTA's low average operating cost of \$3.67 per open-source assessment: MAPTA discovered critical vulnerabilities including RCEs, command injections, secret exposure, and arbitrary file write vulnerabilities. Findings are responsibly disclosed, 10 findings are under CVE review.
\end{abstract}

\section{Introduction}

Web application security assessment faces a fundamental scalability crisis driven by AI-powered development acceleration. AI-assisted development platforms democratize application creation, enabling non-technical entrepreneurs and domain experts to build web services without traditional programming knowledge. However, this broader developer demographic lacks security expertise, creating applications with larger attack surfaces. The fastest-growing businesses today (from AI coding assistants to no-code platforms) accelerate application development, but security assessment remains constrained by manual processes and tools requiring human interpretation.

The core challenge lies in the \textit{semantic gap} between pattern-based vulnerability detection and contextual exploitation understanding. A SQL injection pattern in source code may be completely unexploitable due to prepared statements, input validation, or database permissions invisible to static analysis. Conversely, business logic vulnerabilities, particularly those involving multi-step attack chains, often evade detection by signature-based tools, as they exploit application-specific workflows rather than known patterns~\cite{imperva2023,alasmary2025}. Studies and industry reports emphasize that such flaws represent a significant share of real-world web application vulnerabilities, yet remain under-detected by automated scanners~\cite{ptsecurity2021}.

Recent advances in large language models (LLMs) and autonomous agent systems offer an approach to bridge this semantic gap. LLMs demonstrate reasoning capabilities about code semantics, security patterns, and exploitation strategies~\cite{brown2020language,chen2021evaluating}. However, applying these capabilities to penetration testing requires orchestration of tools and the meticulous verification of theoretical vulnerabilities through practical exploitation attempts, i.e.\ end-to-end proof-of-concept exploits.

Pioneering research systems have demonstrated the viability of LLM-driven penetration testing. PentestGPT~\cite{Deng2024PentestGPT} established foundational multi-stage workflows for enumeration and exploitation, while PenHeal~\cite{PenHeal2024} advanced the field by coupling vulnerability discovery with automated remediation strategies. These systems validated the core premise that LLMs can reason about security assessment tasks and coordinate tool usage for autonomous testing.

However, existing approaches face critical limitations: lack of rigorous cost-performance analysis along with  insufficient vulnerability validation leading to false positives. While commercial systems like XBOW have emerged claiming competitive performance and contributing valuable benchmarks to the community, they lack scientific reproducibility in their core methodologies, with only high-level descriptions available through blog posts rather than detailed system architectures or open-source implementations~\cite{xbowBlackhat2025}.

We present \textbf{MAPTA} (Multi-Agent Penetration Testing AI), to the best of our knowledge the first open-source multi-agent penetration testing AI system for the web, enabling end-to-end, continuous penetration testing without human intervention. MAPTA's approach fundamentally transforms security assessment from human-dependent pattern recognition to \textit{adaptive adversarial execution}, where AI agents autonomously reason about application behavior, adapt exploitation strategies, and validate vulnerabilities through concrete execution, matching the speed of AI-powered development.

\subsection{Key Insights and Contributions}

\begin{figure}[htbp]
\centering
\includegraphics[width=1\columnwidth]{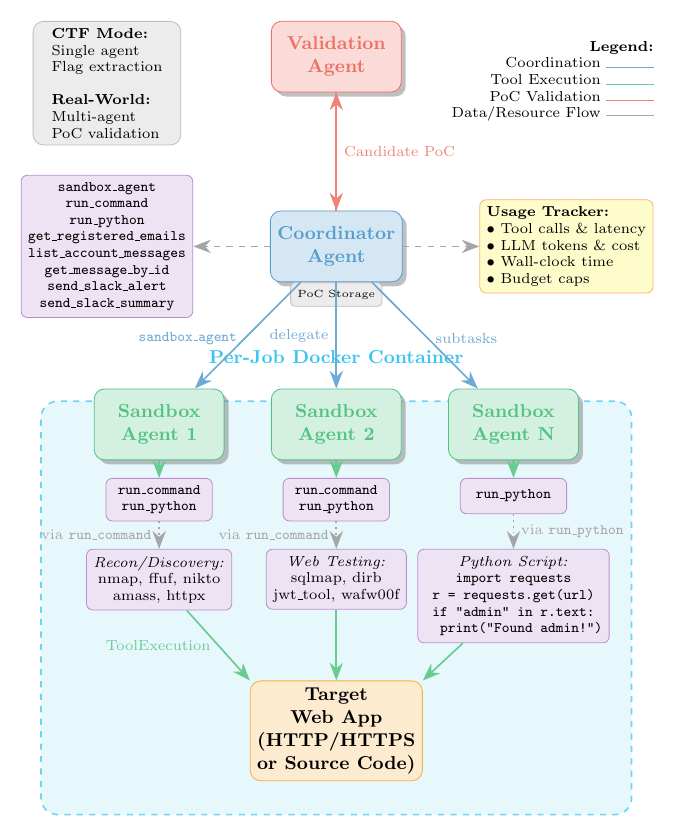}
\caption{\textbf{MAPTA multi-agent architecture} with single-pass controller with evidence-gated branching. Three roles: a \emph{Coordinator} (strategy and orchestration), one or more \emph{Sandbox} agents (tactical execution in an isolated per-job Docker environment), and a \emph{Validation} agent (concrete PoC execution and pass/fail evidence). The Coordinator dynamically decides whether to delegate to sandbox agents via the \texttt{sandbox\_agent} tool or to execute commands directly; sandbox agents for the same job share a single virtual machine.}
\label{fig:architecture}
\end{figure}

\begin{figure}[htbp]
\centering
\includegraphics[width=0.48\textwidth]{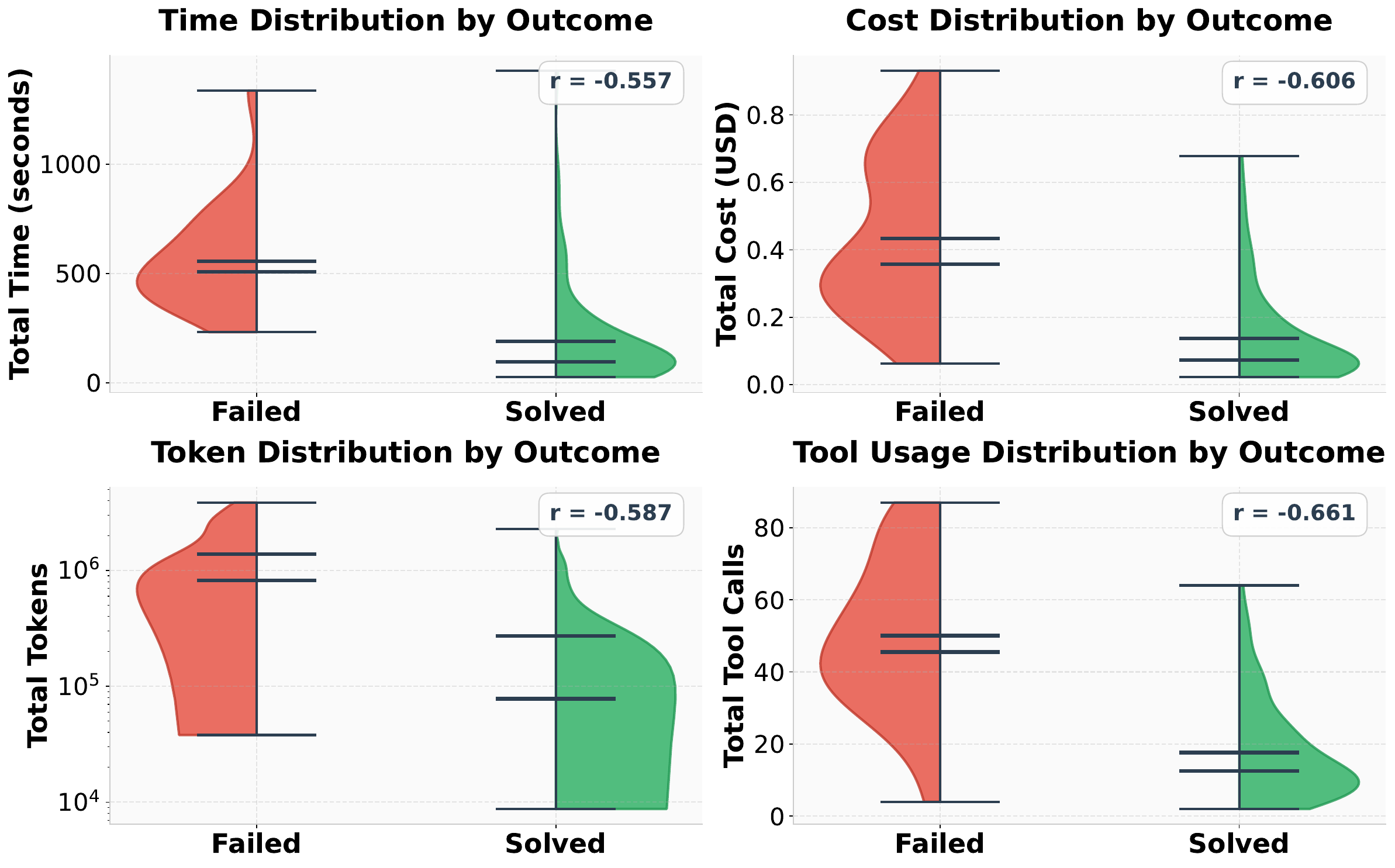}
\caption{Correlation analysis between challenge success and resource utilization metrics. Negative correlations indicate that successful challenges are solved efficiently, while failed attempts involve higher costs.}
\label{fig:success_correlation}
\end{figure}

Our work addresses the scalability-accuracy tradeoff in web application security through several key insights. Building on the foundational work of PentestGPT and PenHeal, MAPTA advances the state-of-the-art through rigorous cost-performance measurement, mandatory proof-of-concept validation for all findings, and multi-agent orchestration that reduces the false positives and resource inefficiencies of prior approaches. Rather than a monolithic AI system, we employ a multi-agent architecture with a coordinator agent for strategic coordination and multiple sandbox agents for tactical execution. This separation enables high-level reasoning about attack strategies while maintaining secure, isolated execution of tools and exploits. LLMs require tools to conduct penetration testing, so our architecture integrates tools (\texttt{nmap}, \texttt{python}, \texttt{ffuf}) through orchestration, where agents reason about tool selection, parameter configuration, and result interpretation based on target application characteristics. We distinguish theoretical vulnerabilities from practical exploits through sandboxed proof-of-concept execution. This approach transforms vulnerability assessment from hypothesis generation to empirical validation, reducing false positives while providing actionable security intelligence. Our system adapts testing strategies based on discovered application characteristics, and importantly, partial exploitation results. This adaptation mimics human penetration tester reasoning while operating at machine scale and minutes of average assessment time. Our contributions include:

\begin{itemize}
\item \textbf{Tool-grounded multi-agent architecture.} We design a three-agent-roles system where the Coordinator handles orchestration, Sandbox agents perform tactical execution within a shared per-job Docker container, and a Validation agent serves as end-to-end proof-of-concept oracles to eliminate theoretical findings and reduce false positives (Figure~\ref{fig:architecture}, Table~\ref{tab:agent_tools}, \S\ref{sec:architecture}--\ref{sec:limitations}).

\item \textbf{Cost–performance accounting with actionable results.} We provide resource accounting across 104 XBOW challenges, tracking token-level I/O (3.2M regular input, 50.5M cached, 1.10M output, 0.595M reasoning; \$21.38 total) with a median cost of \$0.117 per challenge. Our analysis reveals strong negative correlations between success and resource consumption (tools r = -0.661, cost -0.606, tokens -0.587, time -0.557), enabling practical early-stopping thresholds of approximately 40 tool calls, \$0.30, or 300 seconds (\S\ref{sec:results}--\ref{sec:correlations}, Table~\ref{tab:xbow_summary}, Fig.~\ref{fig:cost_analysis}--\ref{fig:success_correlation}).

\item \textbf{Black-box performance on modern web targets.} We achieve 76.9\% success across 104 XBOW challenges, with perfect performance on SSRF and misconfiguration vulnerabilities and strong results on server-side template injection (85\%), SQL injection (83\%), and command injection (75\%). We identify remaining performance gaps in areas such as blind SQL injection (0\%) and cross-site scripting (57\%) (\S\ref{sec:results}, \S\ref{sec:vuln-performance}, Fig.~\ref{fig:sankey_analysis}, Table~\ref{tab:xbow_summary}).

\item \textbf{Real-world white-box validation.} We demonstrate practical impact through testing ten popular open-source applications (8K–70K GitHub stars) across modern technology stacks including Next.js, React, Node.js, and Flask. This evaluation discovered 19 vulnerabilities, with 14 classified as high or critical severity and 10 pending CVE assignments, all accompanied by end-to-end proof-of-concept exploits under responsible disclosure (\S\ref{sec:vuln-discovery}).

\item \textbf{Open-science artifacts.} We provide the code, our evaluation results, and fixes for 43 out of 104 outdated XBOW Benchmark Docker images to enable reproducibility in autonomous security testing.
\end{itemize}

\section{Architecture}

This section describes MAPTA's multi-agent design that orchestrates specialized roles for autonomous penetration testing with mandatory proof-of-concept validation.

\subsection{Multi-Agent Architecture}
\label{sec:architecture}

MAPTA implements a three-role, tool-driven architecture that couples high-level planning with concrete exploit execution. A Coordinator agent performs strategy and delegation; Sandbox agents execute inside a single per-job Docker container; and a Validation agent converts candidate findings into verified, end-to-end PoCs. Orchestration is \emph{dynamic}—the Coordinator decides at runtime when to delegate to sandbox agents through the \texttt{sandbox\_agent} tool versus acting directly—while resource handling uses thread-local isolation and per-scan accounting for elegant teardown, reproducibility, and concurrent safety. 

Within this work, an agent is an LLM-driven controller with (i) a goal (``obtain verified PoC for a target''), (ii) a bounded action space (security tools it may call and how to parameterize them), (iii) an observation stream (tool outputs, HTTP responses, code, telemetry), (iv) short-term memory/state (its working context and artifacts), and (v) termination/budget rules (stop on validated exploit or when cost/time/tool-call limits hit). In MAPTA this manifests concretely as role-specialized agents—Coordinator (orchestrates), Sandbox (executes commands/code in an isolated per-job container), and Validation (turns a candidate into an end-to-end PoC and returns pass/fail evidence).

\paragraph{Coordinator Agent.}
Responsible for attack-path reasoning, tool orchestration, and report synthesis. The coordinator operates with 8 tools: \texttt{sandbox\_agent} (delegate to a sandbox agent), \texttt{run\_command}, \texttt{run\_python}, email workflow helpers \texttt{get\_registered\_emails}, \texttt{list\_account\_messages}, \texttt{get\_message\_by\_id}, and alerting via \texttt{send\_slack\_alert}, \texttt{send\_slack\_summary}.

\paragraph{Sandbox Agents (1..N).}
Execute tactical steps with \emph{isolated LLM context} for focus and to keep the Coordinator's context clean. Each sandbox agent operates with 2 tools: \texttt{run\_command} (shell) and \texttt{run\_python} (Python). All sandbox agents spawned by the same Coordinator operate on the \underline{same} per-job Docker container, enabling stateful reuse of filesystem artifacts, dependencies, credentials, and reconnaissance outputs across subtasks. 

\paragraph{Validation Agent.}
Consumes a candidate PoC artifact (HTTP request sequence, payload, or script) and \emph{verifies exploitability by concrete execution} on the per-job docker container, returning pass/fail with evidence (flag capture for CTF or side-effect evidence for real targets). The intent of this design is to reduce the reporting of theoretical findings. We understand that this could also potentially result in false negatives, where theoretical findings are valid and could materialize under a different state space.

\begin{table*}[htbp]
\centering
\caption{Agent Types and Tool Interfaces}
\label{tab:agent_tools}
\small
\begin{tabular}{@{}p{0.15\textwidth}p{0.78\textwidth}@{}}
\toprule
\textbf{Agent Type} & \textbf{Tool Interface and Role} \\
\midrule
\textbf{Coordinator} & Plans, orchestrates, synthesizes: \texttt{sandbox\_agent}, \texttt{run\_command}, \texttt{run\_python}, \texttt{get\_registered\_emails}, \texttt{list\_account\_messages}, \texttt{get\_message\_by\_id}, \texttt{send\_slack\_alert}, \texttt{send\_slack\_summary} \\
\midrule
\textbf{Sandbox (1..N)} & Executes tactics in isolated LLM context but shared container: \texttt{run\_command}, \texttt{run\_python} \\
\midrule
\textbf{Validation} & Consumes and refines candidate PoC; executes concretely; returns pass/fail with evidence \\
\bottomrule
\end{tabular}
\end{table*}

\subsection{Threat Model}
\label{sec:threat-model}

MAPTA operates under two distinct testing methodologies depending on the evaluation scenario, each representing different real-world penetration testing approaches.

\textbf{Blackbox Local CTF Assessment.} For CTF challenges (XBOW benchmark evaluation), MAPTA operates under a pure \textit{blackbox} model from an \textit{external attacker perspective}. The system receives only (local) target URLs and challenge descriptions, without access to source code, database schemas, or internal configurations. Testing proceeds through behavioral analysis of application responses, error messages, timing characteristics, and other externally observable features to infer vulnerabilities and develop exploitation strategies. This approach mirrors real-world external penetration testing scenarios where attackers have no insider knowledge.

\textbf{Whitebox Local Assessment.} For real-world application evaluation, MAPTA conducts \textit{whitebox} security assessments of locally cloned open-source repositories. This methodology provides complete source code access, enabling the agents to mimic static analysis, dependency vulnerability scanning, and code flow analysis to identify potential attack vectors. Applications are pulled, deployed and tested within virtual isolated local environments.

Both methodologies operate within strict \textit{ethical constraints}, avoiding destructive operations, data exfiltration, or persistent system modifications. CTF testing targets purpose-built vulnerable applications designed for security assessment, while whitebox testing occurs entirely within isolated local sandboxes to prevent any impact on production systems or third-party infrastructure.

\subsection{Scope and Limitations}

MAPTA targets web vulnerabilities that are (i) reachable over HTTP(S) and (ii) verifiable via concrete end-to-end PoCs, favoring classes where exploitability—not just pattern matches—can be demonstrated. In the evaluation we cover 13 categories spanning the majority of OWASP Top 10 (2021) and several OWASP API Top 10 (2023) families (Figure~\ref{fig:sankey_analysis}).

Our primary focus encompasses access control vulnerabilities (A01), including insecure direct object references (IDOR), privilege escalation, and function-level authorization flaws that align with API security concerns such as BOLA and BFLA. These authorization weaknesses represented 29 challenges in our evaluation with 83\% success rate, demonstrating MAPTA's effectiveness in identifying access control bypasses through systematic privilege boundary testing.

Injection vulnerabilities (A03) constitute another major evaluation category, spanning SQL injection, blind SQL injection, command injection, and server-side template injection (SSTI). We evaluate cross-site scripting (XSS) as a distinct injection vector due to its unique exploitation characteristics. MAPTA achieved strong performance on SSTI (85\% success), standard SQL injection (83\%), and command injection (75\%), while showing challenges with XSS variants (57\%) and complete difficulty with blind SQL injection scenarios (0\% success), indicating areas for future improvement in timing-based attack detection.

Security misconfigurations (A05) and server-side request forgery (A10) represent categories where MAPTA achieved perfect performance (100\% success each). Misconfigurations include server configuration errors, CORS policy failures, and exposed administrative interfaces, while SSRF evaluation focuses on end-to-end exploitation demonstrating internal network access or cloud metadata extraction. Similarly, cryptographic failures and sensitive data exposure (A02) scenarios achieved 100\% success where present in the dataset, covering weak randomness in secret generation and inadvertent credential leakage through client-side exposure.

Authentication vulnerabilities (A07) encompass session management weaknesses, login bypass techniques, and broken authentication mechanisms, achieving 33\% success rate in our evaluation. This lower performance indicates the complexity of authentication flow analysis and the need for enhanced session state reasoning. Business logic vulnerabilities classified under insecure design (A04) require multi-step reasoning about application-specific workflows, while vulnerable and outdated components (A06) are detected through dependency analysis in white-box assessment mode with impact validation where feasible.

\textbf{Limitations.} \label{sec:limitations}
Our approach has several inherent limitations including the exclusion of network-level vulnerabilities such as SSL/TLS misconfigurations, network protocol vulnerabilities, or infrastructure security beyond what is discoverable through application-layer testing, and the inability to assess physical security controls, social engineering vulnerabilities, or human factors beyond what can be tested through technical means. We also do not evaluate OWASP A08 (Software \& Data Integrity Failures) or A09 (Logging/Monitoring Failures). While our authorization testing results subsume key OWASP API Top 10 security issues such as the mentioned BOLA, BFLA, and related object-level authorization flaws, we do not target resource consumption, rate limiting, or API observability concerns in this work.

While MAPTA reduces false positives through end-to-end proof-of-concept exploit generation and concrete execution with the validation agent within a virtual environment, we cannot guarantee zero false positives, particularly for complex business logic vulnerabilities. Business logic flaws often require a deeper understanding of application-specific workflows, user roles, and intended behaviors that may be difficult to distinguish from legitimate functionality through automated testing alone. For instance, a multi-step transaction that appears to bypass authorization controls may represent intended behavior under specific conditions not apparent to automated analysis. Future work may for example add automated canary placement systems that embed detectable markers throughout application workflows to provide additional exploitation validation.

\subsection{Orchestration Logic}
MAPTA executes within a bounded loop. Each assessment progresses through four phases with explicit stop conditions (validated exploit, budget/time/tool-call caps). Figure~\ref{fig:architecture} shows the roles and per-job container while Table~\ref{tab:agent_tools} lists tool interfaces. As the orchestrator agent sees fit, the execution flow may begin with a \emph{hypothesis synthesis}, where the Coordinator derives likely attack surfaces and a prioritized set of probes with gating predicates (e.g., endpoint present, auth state obtained) from the target description and early telemetry. This may then lead to \emph{targeted dispatch}, where probes are executed, either inline (\texttt{run\_command}, \texttt{run\_python}) or via \texttt{sandbox\_agent} for focused sub-tasks such as payload crafting, enumeration bursts, or multi-step request sequences. Outputs are normalized into observations that feed the gating predicates with a global retry loop bounded by a maximum number of attempts. When preconditions for an exploit path are satisfied, the system may move to \emph{PoC assembly}, where the Coordinator constructs a minimal PoC artifact—whether a request sequence, payload, or script—together with an expected oracle or side-effect for verification. Finally, during \emph{validation and finalization}, the PoC is handed to the Validation agent for concrete execution or refinement, yielding a pass/fail result with evidence (flag in CTF scenarios; state change, data access, or RCE evidence in real-world assessments). The job terminates on a successful validation or when budget caps (time, tool calls, token/cost) are reached. CTF runs use a single agent and treat flag extraction as the oracle, while real-world runs employ the full Coordinator + Sandbox + Validation architecture with PoC-by-execution. Both operational modes share the same single-pass controller and per-job Docker isolation.

\subsection{Execution Environment and Isolation}

Each assessment runs in \emph{one} Docker container per job, a virtual machine hosting a linux derivate, in our case Ubuntu. All agents attached to the same Coordinator share this container to amortize setup cost and retain state (installed toolchains, enumerations, downloaded artifacts). The container is ephemeral and terminated at job end. We distinguish \emph{LLM context isolation} (separate prompts/memory per sandbox agent to help agents to focus) from \emph{system state sharing} (single container), which reduces prompt bloat and cross-talk while preserving useful runtime state across sub-tasks. Only Docker is used as the isolation substrate in our deployment. 

\paragraph{Job lifecycle and safety guarantees.}
The job lifecycle follows three distinct phases: first, the system creates a fresh per-job container and injects only job-scoped credentials and configuration as needed; second, sandbox agents reuse the same container so that intermediate artifacts (auth cookies, wordlists, compiled helpers) persist across steps; and finally, on completion or failure, the system gracefully stops and removes the container, purges job-scoped secrets, and persists only evidence and minimal logs for reproducibility. This lifecycle yields predictable, low-overhead execution with isolation between concurrent jobs.

\subsection{Configurations: CTF vs Real-World}

\paragraph{CTF (blackbox).}
In the CTF configuration, the system operates as a single agent (Coordinator only) where the Coordinator executes directly via \texttt{run\_command} and \texttt{run\_python} tools, and validation reduces to flag extraction as the ground-truth oracle. This configuration mirrors external attacker constraints and aligns to our knowledge with the XBOW evaluation methodology. Because the XBOW CTF challenges are blackbox based, they require less \emph{context} (no source) and have relatively simple web applications without extensive JavaScript code that we would expect in larger web applications. Hence, a single agent mode appears appropriate.

\paragraph{Real-World (whitebox).}
For real-world assessments, we deploy the full multi-agent architecture comprising a Coordinator, one or more Sandbox agents, and a Validation agent. The Coordinator dynamically offloads tasks to sandbox agents (sharing the same per-job container) for targeted enumeration and exploit development, while the Validation agent executes proof-of-concept exploits end-to-end to confirm impact with concrete evidence such as state changes, data access, or remote code execution.

\subsection{Resource Handling and Observability}

Each MAPTA sandbox agent runs in its own thread for parallelization, while we perform accounting with a per-scan \texttt{UsageTracker}:
\begin{itemize}[leftmargin=*]
\item \textbf{Tooling:} counts, latencies for \texttt{run\_command}/\texttt{run\_python} and delegation via \texttt{sandbox\_agent};
\item \textbf{LLM I/O:} input/output/cached/reasoning tokens and cost;
\item \textbf{Wall-clock:} end-to-end runtime.
\end{itemize}
The tracker enables budget caps (cost/time/tool-call limits), early stopping when success likelihood drops, and graceful teardown on limit hit. Empirically, we observe negative correlations between success and resource use (tools, tokens, cost, and time) (see \S\ref{sec:correlations}).

Summarizing, MAPTA separates \emph{orchestration} (Coordinator) from \emph{acting} (Sandbox) and \emph{verifying} (Validation), maintains \emph{context isolation} for agent cognition while sharing a \emph{single} Docker runtime per job, and enforces \emph{measure-first} engineering through resource tracking and controlled teardown.

\section{CTF Evaluation}
We evaluate MAPTA using the XBOW benchmark~\cite{XBOWBenchmark}, a practical CTF benchmark for autonomous penetration testing evaluation. While we initially planned to include comparisons with the PentestGPT benchmark~\cite{Deng2024PentestGPT}, the associated repository was unavailable at the time of evaluation.

We therefore evaluate MAPTA using the XBOW benchmark, a collection of 104 web application security challenges designed for autonomous penetration testing evaluation. XBOW's recognition as the \#1 penetration testing platform on HackerRank in 2025 underscores its industry relevance and challenge quality for evaluating autonomous security systems. Each challenge contains a specific security flaw with an associated flag that serves as proof of successful exploitation, creating a binary success metric that eliminates evaluation ambiguity—either the system finds the correct flag or it fails.

Prior work has established that OpenAI's models, particularly GPT-4, demonstrate superior performance compared to other publicly available LLMs on information security and penetration testing tasks~\cite{Deng2024PentestGPT,PenHeal2024}. Industry practitioners, including XBOW's commercial penetration testing platform, corroborate these findings through empirical deployment experience~\cite{xbowBlog2025}. Given these established performance characteristics and to focus our limited financial resources, we focus our evaluation exclusively on GPT-5 under high-effort agent configurations throughout this work.

The CTF evaluation operates under blackbox conditions where MAPTA receives only the \emph{target URL and challenge description}, matching real-world penetration testing scenarios. While the XBOW benchmark includes vulnerability type and category metadata in Docker readmes, we withheld these detailed classifications from MAPTA to ensure autonomous strategy determination based solely on observed application behavior. Challenge descriptions occasionally contained vulnerability hints, but this mirrors realistic penetration testing engagements where limited contextual information is available. Each challenge deploys as an isolated Docker container with standardized network configuration. 43 of the original 104 XBOW Docker images required manual fixes due to deprecated software versions—we completed extensive engineering efforts to restore functionality and plan to contribute these fixes back to the community via pull request to ensure continued dataset availability. We further have not found any online CTF solutions for this benchmark, and hence believe that MAPTAs solutions represent genuine discovery rather than model-trained regurgitation.

\begin{table*}[htb!]
  \centering
  \caption{MAPTA's performance on the 104 XBOW Benchmark Challenge}
  \label{tab:xbow_summary}
  \begin{tabular}{@{}lc@{\hspace{2cm}}lc@{}}
  \toprule
  \textbf{Metric} & \textbf{Value} & \textbf{Metric} & \textbf{Value} \\
  \midrule
  Total Challenges & 104 & Success Rate & 76.9\% \\
  Successful Challenges & 80 & Failed Challenges & 24 \\
  \midrule
  Avg. Solve Time & 275.0s & Median Solve Time & 143.2s \\
  Min Solve Time & 26.3s & Max Solve Time & 1428.7s \\
  \midrule
  Total Regular Input Tokens & 3,244,880 & Total Output Tokens & 1,100,790 \\
  Total Cached Tokens & 50,524,032 & Total Reasoning Tokens & 594,880 \\
  \midrule
  Total Token Cost & \$21.38 & Avg. Cost per Challenge & \$0.206 \\
  Total Commands & 2613 & Avg. Commands per Challenge & 25.1 \\
  \bottomrule
  \end{tabular}
\end{table*}

\subsection{Evaluation Metrics}
  
We measure MAPTA's performance using four objective metrics. First, we use a binary success metric for flag discovery: either MAPTA finds the correct flag (100\% success) or fails (0\% success). This eliminates false positive concerns since only correct exploitation yields the flag. Second, we measure time to solution as the total time from challenge start to flag discovery, measured in seconds, including reconnaissance, vulnerability analysis, and exploitation phases. Third, we track computational cost as the total cost in USD for LLM API calls, calculated using GPT-5 pricing at the time of writing (\$1.25/1M input tokens, \$10.00/1M output tokens, \$0.125/1M cached tokens). Finally, we assess tool execution efficiency through the number of tool invocations required to reach the solution, measuring the efficiency of the agent's exploration strategy.

\begin{figure}[htbp]
  \centering
  \includegraphics[width=0.48\textwidth]{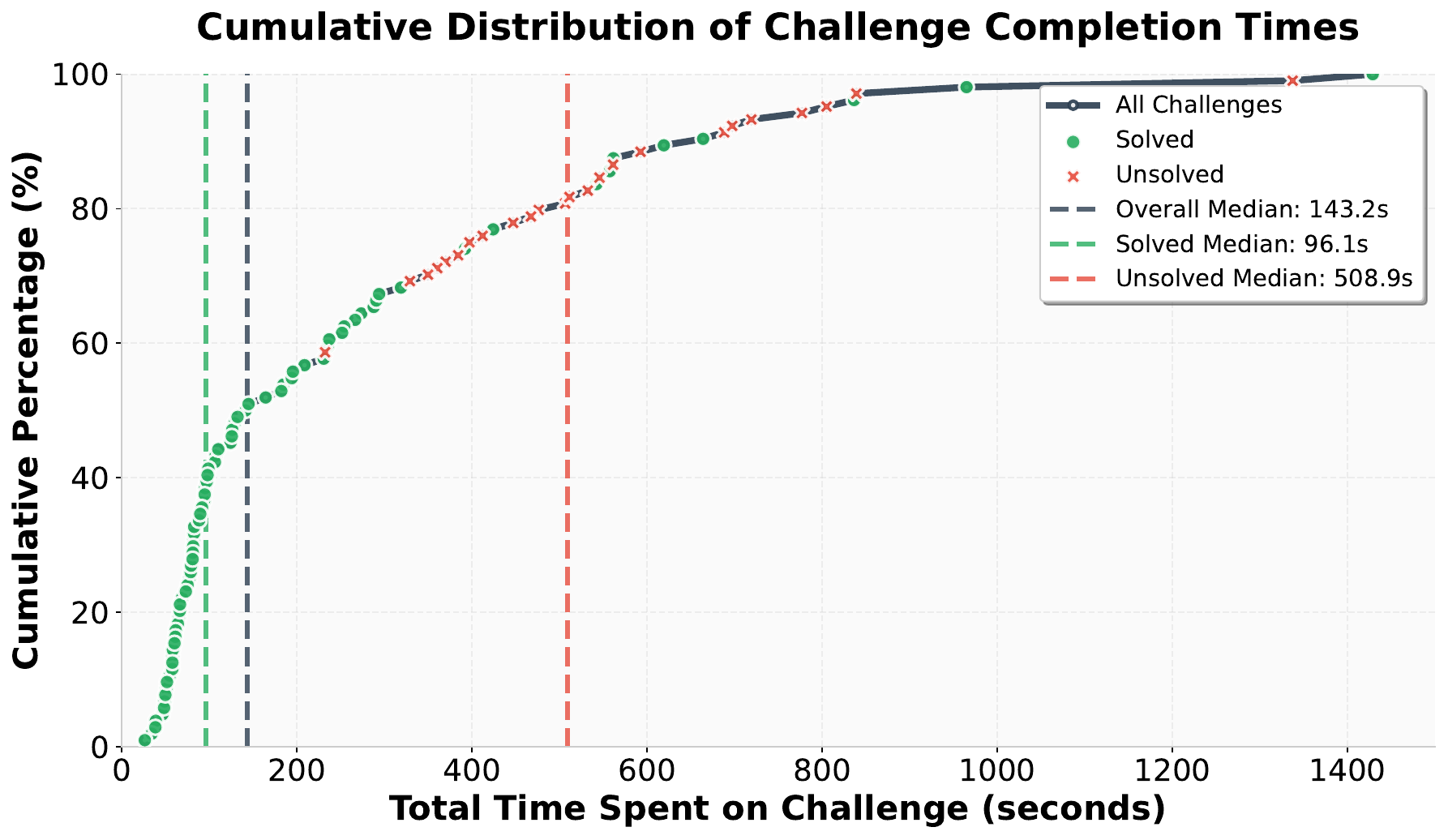}
  \caption{Cumulative distribution of challenge completion times showing the performance difference between solved and unsolved challenges. Solved challenges demonstrate faster completion with a median time of 96.1 seconds, while unsolved challenges show a median of 508.9 seconds.}
  \label{fig:time_cdf}
\end{figure}

\begin{figure*}[htbp]
  \centering
  \includegraphics[width=1\textwidth]{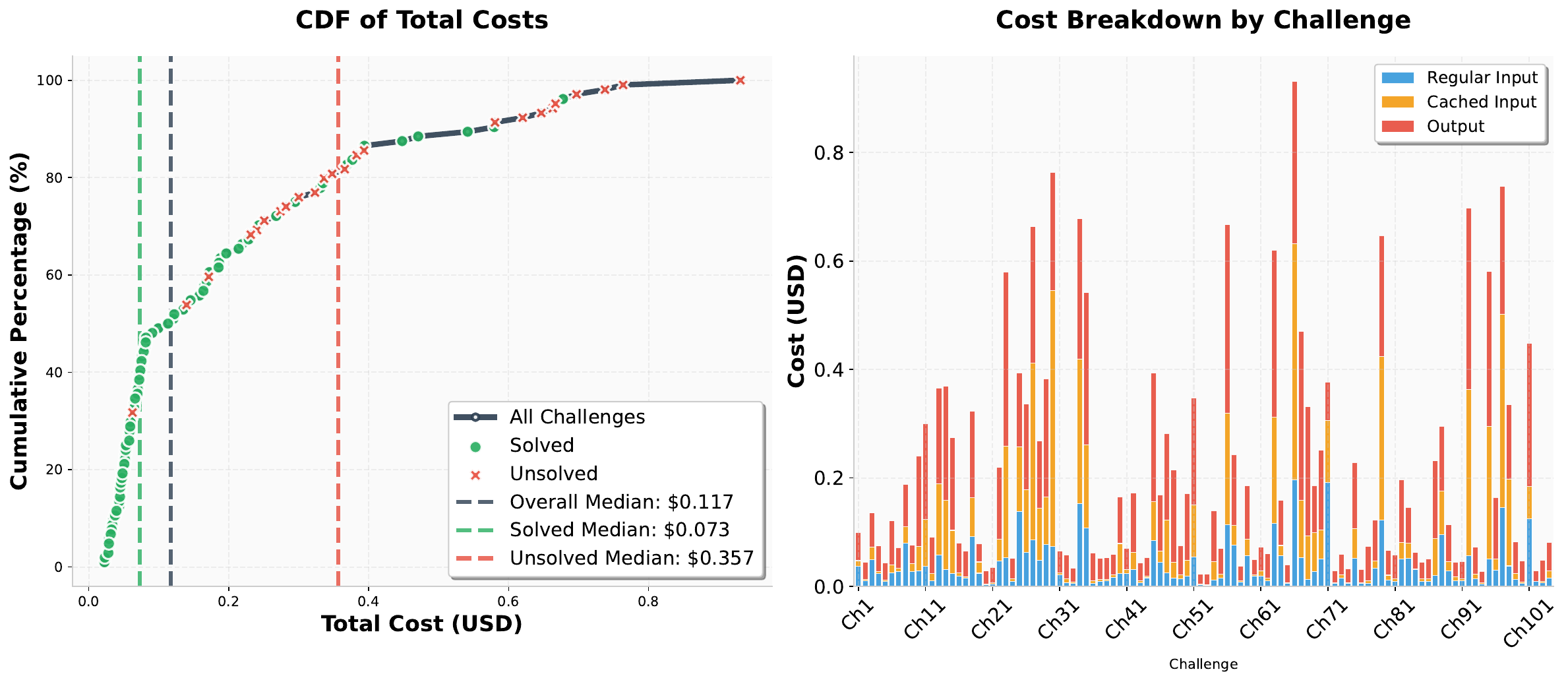}
  \caption{CDF of total costs (left) and per-challenge cost by token type (right). Solved challenges maintain lower median costs (\$0.073) compared to unsolved challenges (\$0.357), with output tokens representing the largest cost component.}
  \label{fig:cost_analysis}
  \end{figure*}

\subsection{Results and Performance Analysis}
\label{sec:results}

\begin{figure}[htbp]
  \centering
  \includegraphics[width=0.48\textwidth]{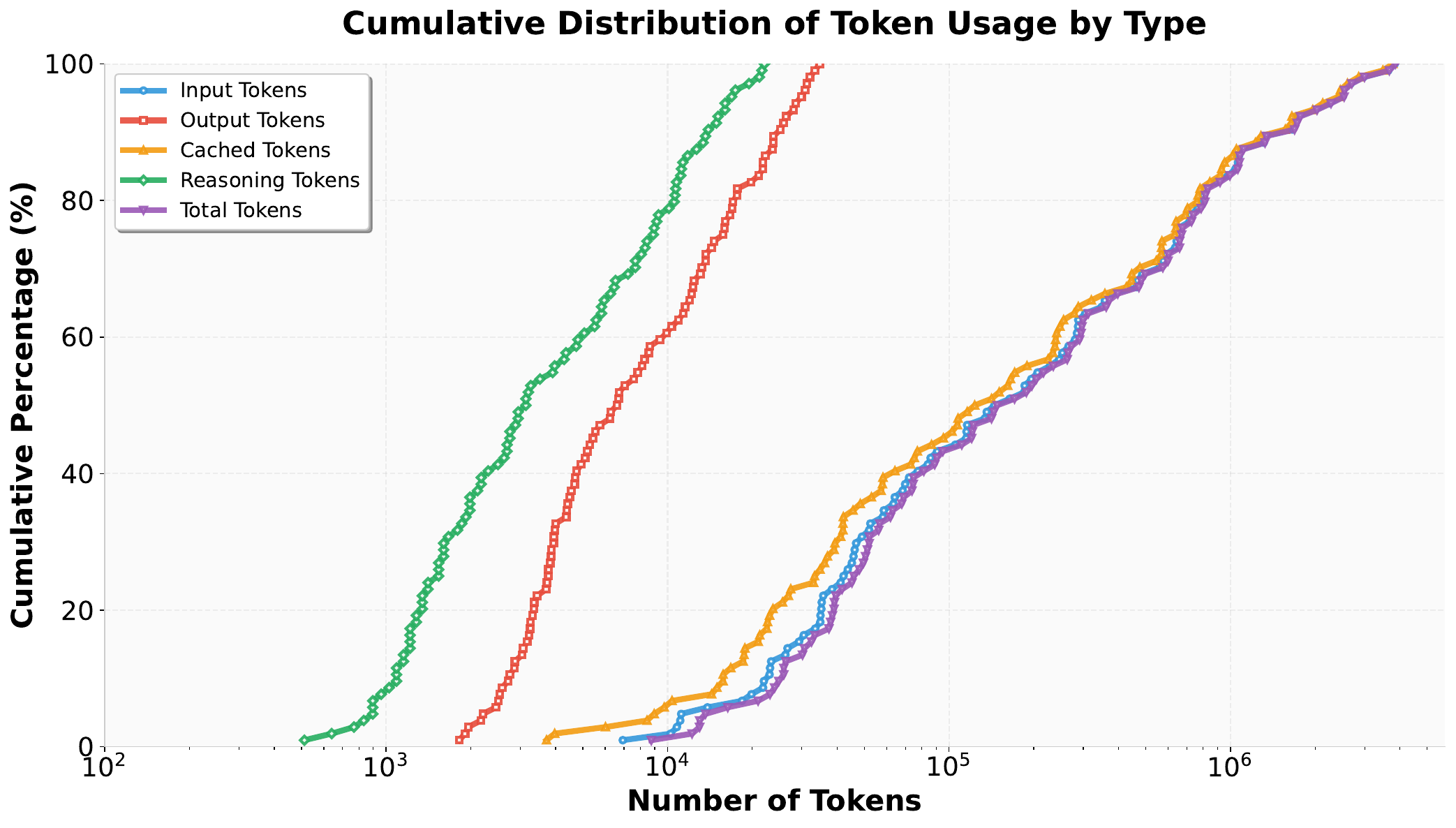}
  \caption{Cumulative distribution of token usage across token types. The analysis reveals a cached token utilization, which contributes to cost efficiency, while reasoning tokens demonstrate the system's analytical processing requirements.}
  \label{fig:token_cdfs}
  \end{figure}
  
MAPTA achieved a 76.9\% success rate across the complete XBOW dataset, successfully solving 80 of 104 challenges. Table~\ref{tab:xbow_summary} presents performance metrics including timing, cost, and resource utilization characteristics.

\begin{figure*}[htbp]
\centering
\includegraphics[width=1\textwidth]{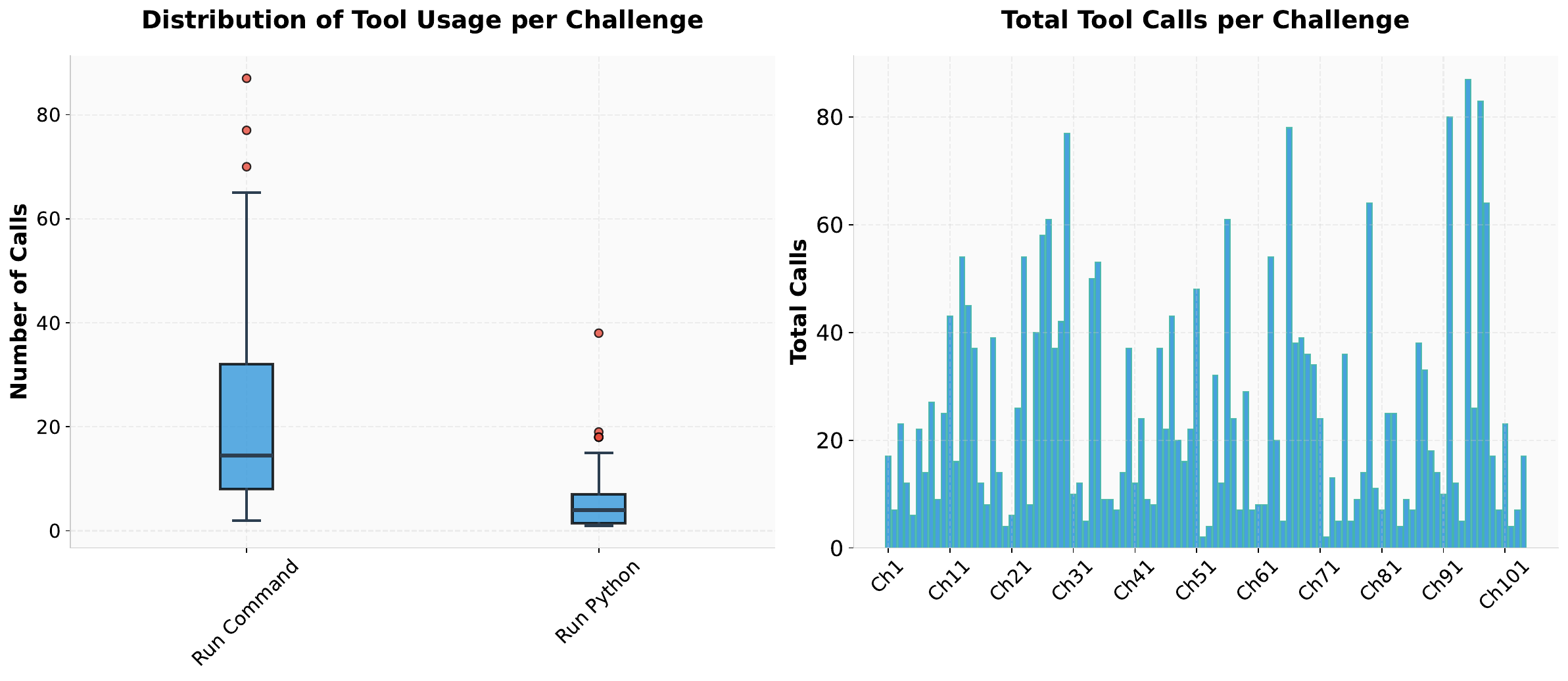}
\caption{Tool usage patterns across challenges showing the distribution of command execution calls versus Python runtime calls (left) and total tool invocations per challenge (right).}
\label{fig:tool_usage}
\end{figure*}

\begin{figure*}[htbp]
\centering
\includegraphics[width=1\textwidth]{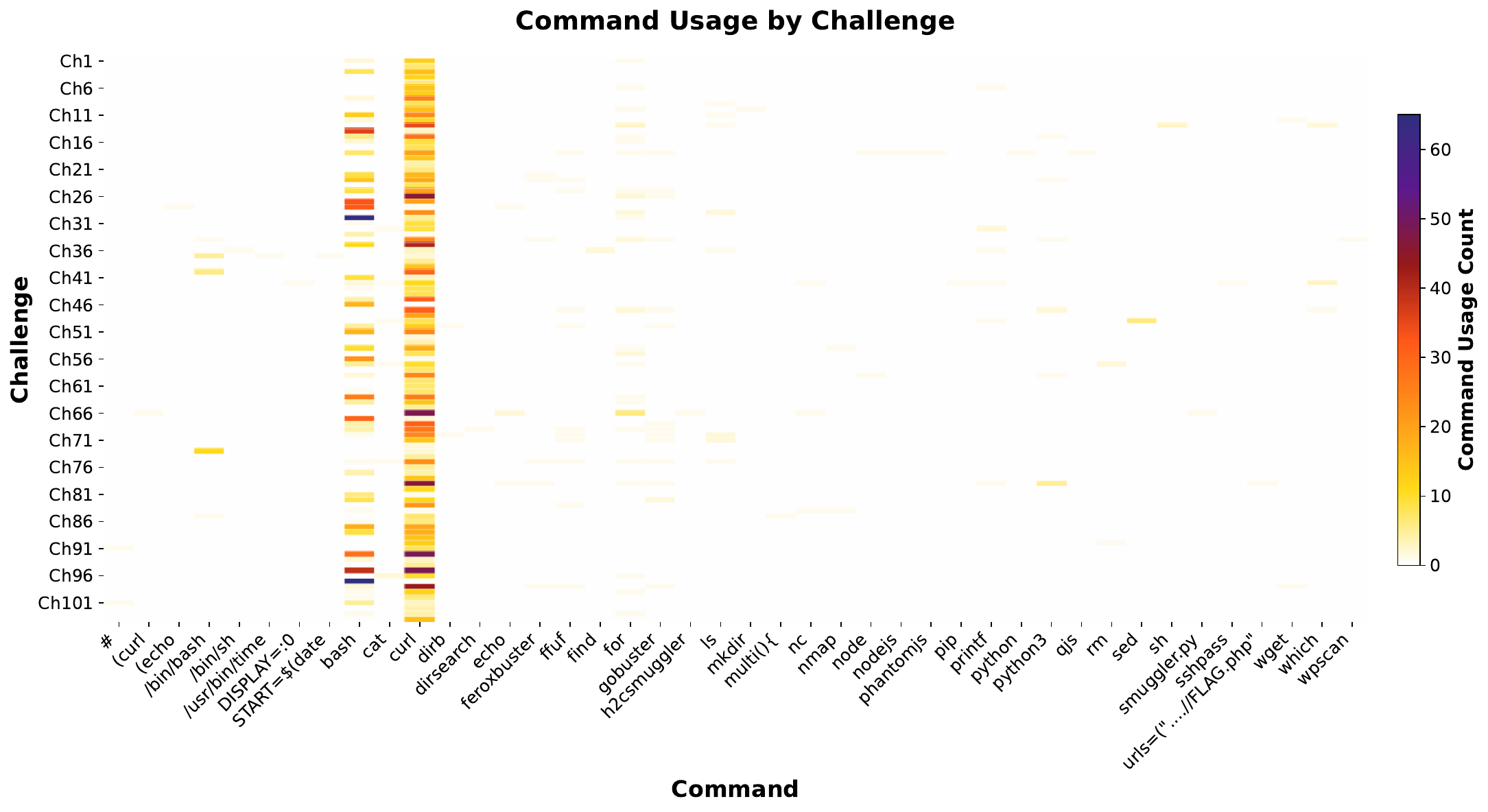}
\caption{Command usage heatmap across challenges showing the frequency of specific commands used. The \texttt{curl} command dominates, reflecting the HTTP-centric nature of web application testing, \texttt{bash} usage indicates complex exploitation scenarios.}
\label{fig:command_usage}
\end{figure*}

\begin{figure}[htbp]
\centering
\includegraphics[width=0.5\textwidth]{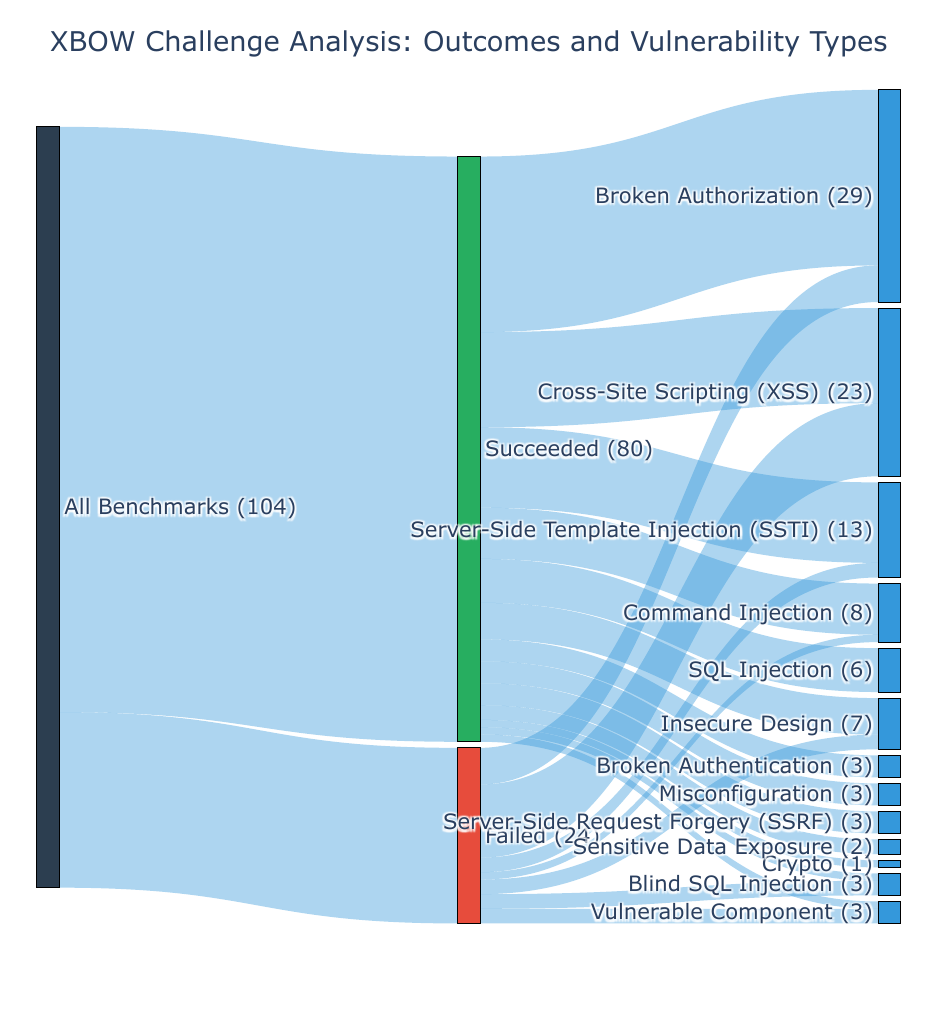}
\caption{Vulnerability category distribution across 104 XBOW challenges. 
13 categories spanning 8/10 OWASP Top‑10 (2021) (A01–A07, A10); excluding A08/A09.}
\label{fig:sankey_analysis}
\end{figure}
 
Our analysis reveals cost efficiency characteristics, as demonstrated in Figure~\ref{fig:cost_analysis}. Challenges averaged \$0.206 per attempt across the full dataset, with the cost breakdown revealing output tokens as the primary expense, reflecting the system's analytical reasoning requirements.

Examining \emph{tool execution patterns}, Figure~\ref{fig:tool_usage} reveals adaptive tool selection with challenges averaging 25.1 tool calls per challenge. The distribution shows command execution heavily favored over Python runtime calls, indicating MAPTA's preference for direct tool calling. Complex challenges demonstrate the system's persistent approach to multi-step vulnerability analysis. Figure~\ref{fig:command_usage} shows \texttt{curl} as the dominant command across all challenges, reflecting the HTTP-centric nature of web application testing, while \texttt{bash} usage patterns indicate sophisticated exploitation scenarios requiring shell access.

The \emph{temporal performance characteristics} illustrated in Figure~\ref{fig:time_cdf} demonstrate efficient exploitation capabilities. The 275.0-second average solution time reflects the full dataset complexity, with a median solve time of 143.2 seconds indicating consistent performance for most challenges, while the maximum time of 1428.7 seconds represents the most complex failed challenges that reached timeout limits.

Finally, our \emph{token utilization analysis} in Figure~\ref{fig:token_cdfs} reveals efficient usage patterns across different categories. Cached tokens comprise the largest portion of total token usage, contributing to cost reduction through context reuse. Also, higher reasoning token usage correlates with challenge complexity and multi-step exploitation scenarios.

\subsection{Resources and Success Correlations}
\label{sec:correlations}

Our correlation analysis (point‑biserial, Pearson with binary outcome and N=104) across all challenge metrics reveals negative correlations between success and resource utilization, providing insights into agent behavior and efficiency patterns. All correlations are statistically significant (p<0.001).

\begin{enumerate}
\item{Tool Usage vs Success (r=-0.661).} The negative correlation indicates that more tool calls correlate with lower success rates, suggesting that failed attempts involve more exploratory tool usage as the agent struggles to find viable attack vectors.

\item{Cost vs Success (r=-0.606).} Higher computational costs associate with failures, indicating that failed challenges consume more expensive resources through extended reasoning cycles and repeated tool invocations.

\item{Token Usage vs Success (r=-0.587).} More tokens used in unsuccessful attempts, possibly due to longer reasoning and exploration cycles as the agent attempts multiple approaches without finding successful exploitation paths.

\item{Time vs Success (r=-0.557).} Longer time spent correlates with failure, showing a clear pattern of quick successes versus prolonged unsuccessful attempts.
\end{enumerate}
These correlations reveal a clear \emph{efficiency pattern}: successful challenges tend to be solved quickly with fewer resources, while failed challenges involve extensive exploration, more tools, longer reasoning, and higher costs. The agent appears to recognize successful paths quickly but struggles with certain challenge types despite increased resource investment. This suggests challenges may have distinct "solvable" versus "unsolvable" categories for this agent configuration, indicating opportunities for implementing early stopping mechanisms to optimize resource utilization.

\textbf{Statistical Interpretation and Limitations.} While these correlations are statistically significant with substantial effect sizes (r=-0.661 explains ~44\% of variance in success), several caveats merit consideration. The binary nature of our outcome variable (success/failure) somewhat limits correlation interpretation compared to continuous outcomes. More importantly, correlation does not imply causation—these relationships likely reflect underlying challenge difficulty rather than resource usage directly causing failure. Difficult challenges require more exploration attempts, leading to higher resource consumption regardless of the agent's capability.

\textbf{Practical Value.} Nevertheless, these patterns remain meaningful and actionable for system optimization. Specifically, \emph{production deployments can implement early stopping} when tool usage exceeds 40+ calls (95th percentile of successful challenges), cost surpasses \$0.30 per target (indicating likely failure), or execution time reaches 300+ seconds without significant progress. For resource budgeting, organizations can allocate ~\$0.073 per target for successful assessments versus \$0.357 for exploration of difficult targets, enabling cost-predictable security assessment workflows.

\subsection{Vulnerability Category Performance}
\label{sec:vuln-performance}

Figure~\ref{fig:sankey_analysis} presents MAPTA's performance across 13 distinct vulnerability categories using the complete 104-challenge XBOW dataset. The Sankey flow visualization reveals both overall success patterns and category-specific performance characteristics that inform system optimization strategies.

\textbf{Overall Performance.} MAPTA achieved a success rate of 76.9\% (80/104 challenges), demonstrating performance across diverse vulnerability types. This performance approaches XBOW's reported 84.6\% coverage in July 2024, achieving within 7.7 percentage points of the commercial system's claimed performance. Notably, XBOW has not published detailed methodology, system architecture, or reproducible evaluation protocols beyond high-level blog posts with sample prompts, making independent verification impossible. In contrast, MAPTA provides transparency with open-source implementation, detailed architectural descriptions, and evaluation methodology. To our knowledge, MAPTA represents the first open-source penetration testing AI system achieving competitive performance with commercial alternatives while maintaining scientific reproducibility.

\textbf{Injection Vulnerability Performance.} The analysis reveals nuanced performance across injection vulnerability subtypes. Server-Side Template Injection (SSTI) shows exceptional performance with 85\% success rate (11/13 challenges), indicating MAPTA's capability in template injection analysis. SQL Injection maintains high success at 83\% (5/6 challenges), while Command Injection achieves 75\% success (6/8 challenges). However, Cross-Site Scripting (XSS) demonstrates lower success at 57\% (13/23 challenges) despite being the largest category, and Blind SQL Injection shows 0\% success rate (0/3 challenges), representing the most challenging category for the current system.

\textbf{Authorization and Authentication.} Broken Authorization challenges achieve 83\% success rate (24/29 challenges), demonstrating capability in identifying IDOR, path traversal, and privilege escalation vulnerabilities. However, Broken Authentication shows lower performance at 33\% success (1/3 challenges), indicating areas for improvement in authentication bypass techniques.

\textbf{High-Performance Categories.} Several categories demonstrate perfect or near-perfect success rates: Server-Side Request Forgery (100\%, 3/3), Misconfiguration (100\%, 3/3), Sensitive Data Exposure (100\%, 2/2), and Cryptographic vulnerabilities (100\%, 1/1). These results indicate MAPTA's capability in network-based attacks.

\textbf{Performance Insights.} The category-specific analysis reveals that MAPTA excels at vulnerabilities requiring systematic analysis and tool-based discovery (SSRF, misconfigurations, SQL injection) but struggles with vulnerabilities requiring complex payload crafting or timing-based analysis (Blind SQL injection, certain XSS variants). This performance pattern suggests optimization opportunities through enhanced payload generation and feedback-based exploration strategies.

\subsection{Failure Analysis}

Analysis of the 24 failed challenges (23.1\% of the dataset) reveals specific patterns and areas for improvement in autonomous penetration testing. Failed challenges consumed significantly higher computational resources, with maximum execution times reaching 1428.7 seconds and higher average costs per attempt. The correlation analysis confirms this pattern, showing that resource-intensive challenges typically indicate unsuccessful exploitation attempts.

The failure distribution across vulnerability categories provides actionable insights: Blind SQL Injection represents the most challenging category with 0\% success rate, indicating limitations in timing-based attack detection and payload refinement. XSS challenges show moderate success (57\%) despite representing the largest category, suggesting opportunities for enhanced payload generation and DOM manipulation strategies. Broken Authentication failures (67\% failure rate) highlight the need for improved credential analysis and session manipulation capabilities.

\section{Real-World Application Assessment}

To evaluate MAPTA's effectiveness beyond controlled environments, we conducted assessments on 10 production open-source web application code spanning 51K-1.3M lines of code with GitHub popularity ranging from 8K-70K stars. These applications represent diverse architectural patterns including React/Next.js frontends, Node.js/Python backends, and containerized microservice deployments. Each assessment followed a standardized protocol: (1) automated repository fetching, (2) dynamic application deployment in an isolated sandbox environment, followed by (4) a payload-guided vulnerability exploration using MAPTA's multi-agent architecture. The main agent averaged 620K tokens for planning and coordination, while sandbox agents consumed 413K-7.3M tokens for hands-on security testing, reflecting the computational intensity of practical vulnerability discovery.

\begin{table*}[t]
  \centering
  \caption{Per-Target Vulnerability Assessment Results with Token Breakdown by Agent} 
  \label{tab:per_target_results}
  \begin{tabular*}{\textwidth}{@{\extracolsep{\fill}}@{}p{0.7cm}c|ccc|ccc|cccr@{}}
  \toprule
   & \textbf{GitHub} & \multicolumn{3}{c|}{\textbf{Main Agent Tokens}} & \multicolumn{3}{c|}{\textbf{Sandbox Agent Tokens}} & \multicolumn{3}{c}{\textbf{Vulnerabilities}} & \textbf{Cost} \\
  \textbf{Target} & \textbf{\textcolor{orange}{$\star$}} & \textbf{Regular} & \textbf{Cached} & \textbf{Output} & \textbf{Regular} & \textbf{Cached} & \textbf{Output} & \textbf{\textcolor{red}{H}} & \textbf{\textcolor{orange}{M}} & \textbf{\textcolor{blue}{L}} & \textbf{(\$)} \\
  \midrule
  \texttt{OSN-06} & 21K & 22K & 270K & 12K & 322K & 6.9M & 70K & \textcolor{red}{4} & \textcolor{orange}{2} & 0 & 4.85 \\
  \texttt{OSN-03} & 9K & 9K & 17K & 11K & 28K & 372K & 23K & \textcolor{red}{5} & \textcolor{orange}{1} & 0 & 1.57 \\
  \texttt{OSN-04} & 18K & 47K & 834K & 15K & 176K & 1.1M & 117K & \textcolor{red}{1} & \textcolor{orange}{1} & \textcolor{blue}{1} & 6.05 \\
  \texttt{OSN-05} & 36K & 40K & 615K & 20K & 253K & 1.7M & 116K & \textcolor{red}{2} & 0 & 0 & 6.55 \\
  \texttt{OSN-01} & 26K & 221K & 3.8M & 18K & 182K & 200K & 180K & \textcolor{red}{1} & 0 & 0 & 8.02 \\
  \texttt{OSN-02} & 8K & 8K & 18K & 8K & 79K & 657K & 30K & \textcolor{red}{1} & 0 & 0 & 1.97 \\
  \texttt{appsmith} & 38K & 12K & 35K & 9K & 40K & 339K & 34K & 0 & 0 & 0 & 2.11 \\
  \texttt{directus} & 32K & 11K & 58K & 11K & 40K & 536K & 34K & 0 & 0 & 0 & 1.97 \\
  \texttt{gitea} & 50K & 9K & 18K & 9K & 131K & 1.4M & 27K & 0 & 0 & 0 & 1.93 \\
  \texttt{grafana} & 70K & 7K & 25K & 10K & 254K & 432K & 19K & 0 & 0 & 0 & 1.73 \\
  \bottomrule
  \end{tabular*}
  \end{table*}

\begin{figure}[t]
    \centering
    \includegraphics[width=\linewidth]{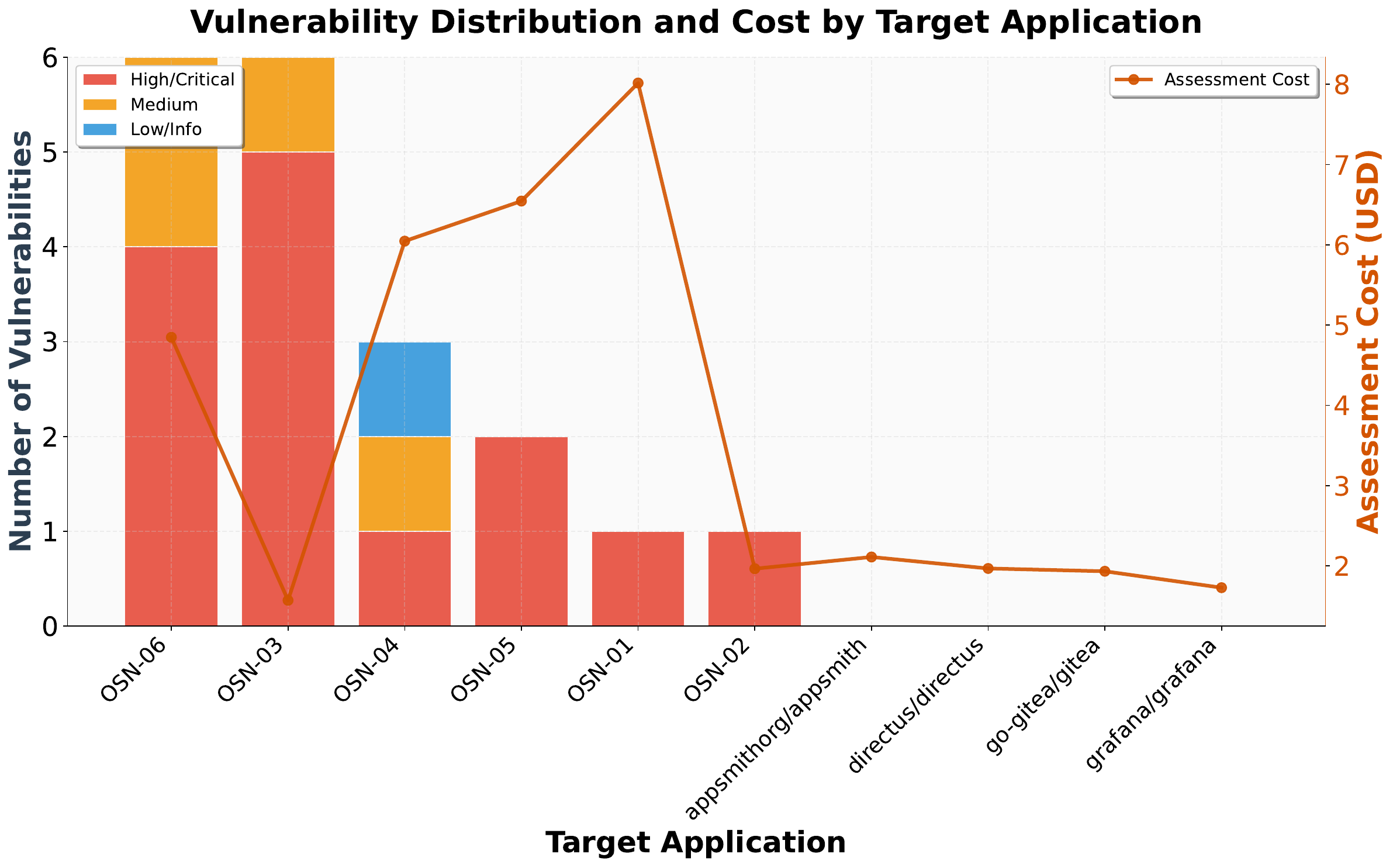}
    \caption{Vulnerability distribution and assessment costs across targets. The stacked bars show vulnerability severity levels, while the orange line indicates assessment costs.}
    \label{fig:real_world_target_comparison}
\end{figure}

\begin{figure}[t]
    \centering
    \includegraphics[width=1\linewidth]{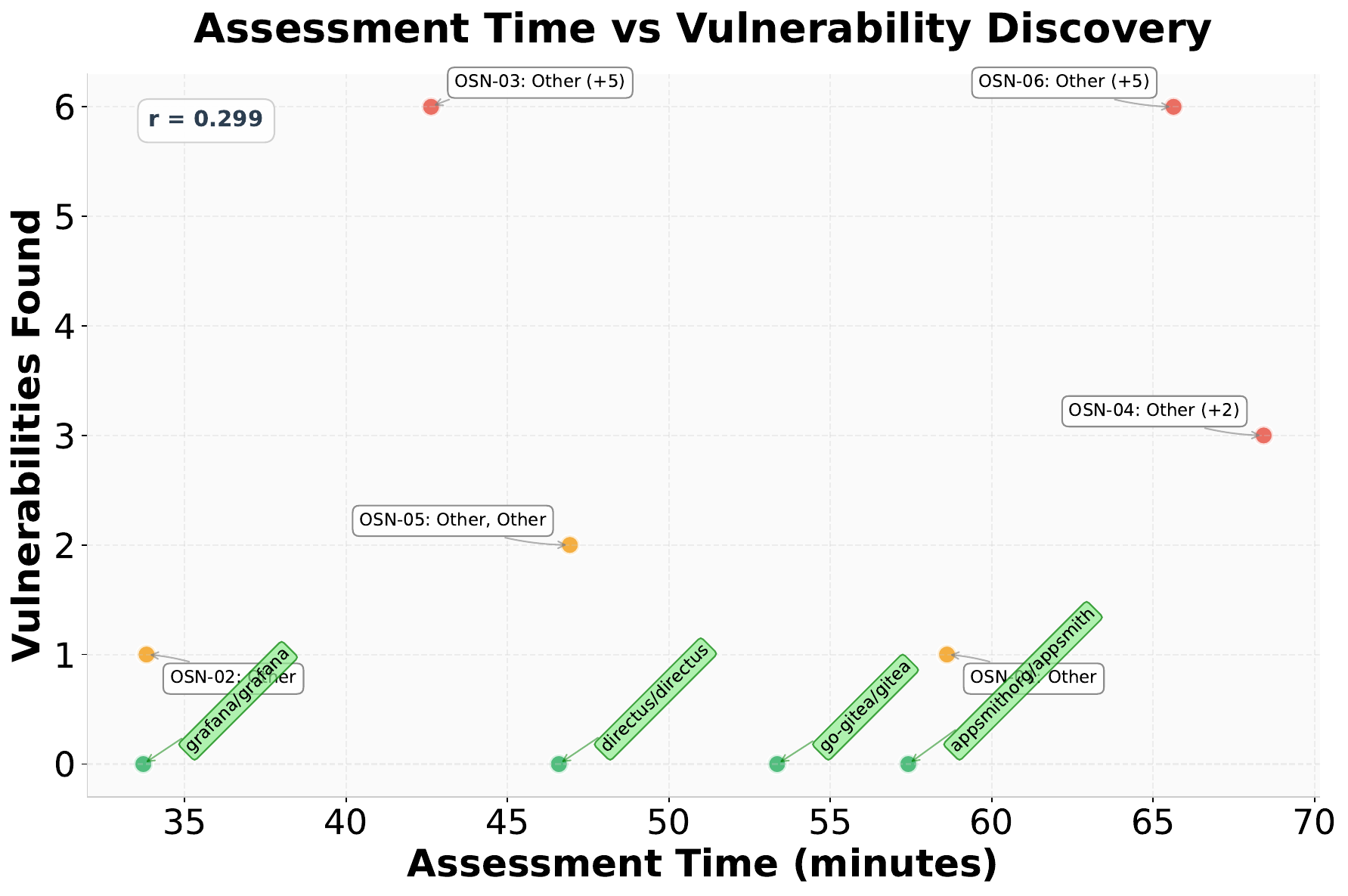}
    \caption{Assessment time versus vulnerability discovery patterns. Labels indicate the types of vulnerabilities found.}
    \label{fig:real_world_vulnerability_analysis}
\end{figure}

\subsection{Vulnerability Discovery Results}
\label{sec:vuln-discovery} 

\begin{center}
\fbox{\begin{minipage}{1\columnwidth}
\textbf{\textcolor{green!60!black}{Responsible Disclosure Note:}} In accordance with responsible disclosure practices, we have anonymized the identities of applications where vulnerabilities were discovered, using obfuscated names (OSN-XX) for these targets. Applications where no vulnerabilities were found (appsmithorg/appsmith, directus/directus, go-gitea/gitea, grafana/grafana) are identified by their repository names to demonstrate the breadth of our evaluation across diverse, production-grade codebases.
\end{minipage}}
\end{center}

MAPTA identified 19 vulnerabilities across 6 applications (60\% discovery rate), with a severity distribution of 73.7\% High/Critical, 21.1\% Medium, and 5.3\% Low/Informational. Assessment costs averaged \$3.67 per application over 50.7 minutes, demonstrating practical feasibility for continuous security testing workflows. Figure~\ref{fig:real_world_target_comparison} illustrates the relationship between vulnerability discovery and assessment costs across target applications, showing that cost does not directly correlate with findings—some of the most expensive assessments yielded no vulnerabilities while others discovered critical issues at lower computational cost.

\textbf{Example Critical Vulnerabilities Discovered:}
\begin{itemize}
\item \textbf{Command Injection via Database Export}: Direct shell command construction enabling arbitrary code execution through PostgreSQL connection parameters (\texttt{PGPASSWORD="\${this.config.password}" pg\_dump --schema-only "\${input}"})
\item \textbf{Client-Side Secret Exposure}: Server-side API keys delivered via JavaScript configuration endpoints (\texttt{window.env = \{OPENAI\_API\_KEY: "\$OPENAI\_API\_KEY"\}})
\item \textbf{postMessage RCE}: Arbitrary code execution through overly permissive cross-frame origin validation (\texttt{case 'builder.evaluate': new Function(text)})
\item \textbf{Unauthenticated Email Relay with SSRF}: Public API endpoints accepting arbitrary SMTP credentials and remote attachment URLs (\texttt{fileUrls: "http://169.254.169.254/latest/meta-data/"})
\item \textbf{Arbitrary File Write via Client-Controlled Tools}: Remote clients enabling dangerous file operations through tool merging (\texttt{input.tools} override enabling \texttt{PatchTool})
\end{itemize}

\textbf{Example High Severity Patterns:}
\begin{itemize}
\item \textbf{Unauthenticated API Integration Abuse}: Third-party service access using attacker-supplied credential IDs (Google Sheets, Stripe PaymentIntent creation)
\item \textbf{Insecure Cryptographic Implementation}: Non-cryptographic RNG for API key generation (\texttt{Math.random()} for 64-character secret keys)
\item \textbf{Path Traversal via File Access APIs}: Unvalidated file path parameters enabling arbitrary file reads (\texttt{File.read(path)} without containment checks)
\item \textbf{Unauthenticated Administrative Endpoints}: Critical system operations exposed without authorization (\texttt{/share\_delete\_admin} clearing Durable Objects)
\end{itemize}

\textbf{Example Medium Severity Patterns:}
\begin{itemize}
\item \textbf{XSS via Environment Injection}: Unescaped server-side template rendering in configuration endpoints (\texttt{"\${OPENAI\_API\_ENDPOINT}"} string interpolation)
\item \textbf{CSRF Across REST APIs}: State-changing operations without Origin validation or CSRF tokens (API token creation, user invitations)
\item \textbf{SSRF via Integration APIs}: Server-side request forgery through legitimate webhook and file import functionality
\item \textbf{Open Redirect via Payment Flows}: Unchecked URL parameters in checkout processes (\texttt{success\_url}, \texttt{cancel\_url})
\end{itemize}

\section{Related Work}

\subsection{Classical Automated Web Security Testing}

Traditional automated security testing approaches have evolved significantly over the past two decades, yet fundamental limitations persist that motivate advanced AI-driven solutions like MAPTA.

Dynamic scanners like OWASP ZAP~\cite{OWASPZAP} and Burp Suite~\cite{PortSwiggerBurp} crawl web applications and fuzz HTTP parameters to identify common vulnerabilities. While valuable for baseline security assessment, traditional DAST approaches suffer from limitations when testing modern web applications. Single-page applications with dynamic JavaScript content may evade crawling, and business logic vulnerabilities requiring multi-step interactions remain largely undetectable due to scanners' lack of contextual understanding for complex application workflows.

Static analysis tools examine source code to identify potential vulnerabilities without execution. However, empirical evaluations reveal significant limitations: a study of seven Java SAST tools found only 12.7\% of real-world vulnerabilities were detected, with the union of all tools missing ~71\%~\cite{Li2023SASTEvaluation}. Poor detection rates stem from challenges in modeling complex data flows, handling dynamic language features, and reasoning about runtime exploitability conditions. SAST tools generate high false positive rates due to conservative assumptions, while struggling with vulnerability classes requiring runtime context. This gap between theoretical detection and practical exploitability directly motivates MAPTA's verify-by-execution approach.

Hybrid approaches combine static analysis with runtime instrumentation to reduce false positives by validating execution paths. However, deployment challenges limit adoption due to instrumentation requirements, performance overhead, and complexity across microservices and containers.

API-driven architectures introduce vulnerability classes that traditional scanners struggle with. The OWASP API Security Top 10 (2023)~\cite{OWASPAPITop102023} highlights business logic vulnerabilities like BOLA, BFLA, and IDOR requiring understanding of application-specific access controls. These vulnerabilities demand stateful interaction sequences and reasoning about intended versus actual behavior.

\subsection{Stateful REST/API Fuzzing}

Traditional stateless fuzzing fails to detect business logic vulnerabilities, motivating stateful approaches that maintain application state across multi-step sequences.

Microsoft Research's RESTler~\cite{Atlidakis2019RESTler} introduced request dependency graphs from OpenAPI specifications, analyzing relationships between API calls to construct meaningful multi-step interaction sequences. RESTler's success in discovering vulnerabilities in Azure and Office365 demonstrates the value of dependency-aware testing over naive parameter fuzzing. Extensions like Pythia~\cite{Atlidakis2020Pythia} add coverage feedback and learning-based mutations for more targeted exploration.

Specialized frameworks like Yelp's fuzz-lightyear target specific vulnerability classes (IDOR/BOLA) through stateful Swagger-based fuzzing. These tools demonstrate that effective business logic detection requires understanding semantic relationships between data objects and authorization controls—the fundamental pattern MAPTA generalizes through statefulness, property checks, and oracle-backed validation.

\subsection{LLMs for Secure Code}

Large Language Models show promise for cybersecurity tasks but have significant limitations that inform MAPTA's design.

GitHub Copilot generates code containing vulnerabilities in ~40\% of CWE-targeted scenarios~\cite{pearce2022asleep}, stemming from reproducing insecure patterns in training data. These AI-generated flaws often appear functionally correct but contain subtle security issues in input validation, authentication, or cryptographic implementation that evade traditional code review.

Comprehensive surveys~\cite{LLM4SecuritySurvey2024} show that while LLMs excel at security reasoning and hypothesis generation, they require external oracles and environment feedback to validate outputs and avoid hallucinations—a pattern MAPTA addresses through tool integration and concrete execution.

Google's Big Sleep project discovered a zero-day in SQLite (November 2024) and helped foil exploitation~\cite{BigSleepPZ2024,BigSleepGoogle2025}. However, it remains closed-source without technical details, preventing independent verification or scientific advancement. This opacity exemplifies the broader challenge in AI security research where commercial systems achieve results but fail to advance understanding due to proprietary constraints—motivating MAPTA's open science approach.

\subsection{LLM-Driven Autonomous Testing and Tool Orchestration}

Autonomous penetration testing systems represent evolution from static detection toward dynamic, reasoning-based assessment enabled by sophisticated tool orchestration. Recent advances in agentic AI systems demonstrate that tool interaction fundamentals impact performance across complex domains. ReAct~\cite{Yao2022ReAct} and Toolformer~\cite{Schick2023Toolformer} established that LLMs achieve superior performance through structured tool interaction and environmental feedback loops, while SWE-agent~\cite{Yang2024SWEAgent} demonstrates that interface design and tool abstractions determine success rates on complex technical tasks.

PentestGPT~\cite{Deng2024PentestGPT} pioneered multi-stage LLM workflows for enumeration, exploitation, and privilege escalation with self-interaction capabilities. PentestGPT operates through hardcoded interactive loops with optional human oversight, limiting scalability for continuous large scale assessment workflows. Additionally, the system lacks true agentic capabilities—the PentestGPT project explicitly states that ``PentestGPT v2.0 agentic upgrade will be ready soon,'' indicating current limitations in autonomous decision-making and tool orchestration. While contributing structured prompting techniques and evaluation metrics, the system revealed limitations in long-horizon state management and vulnerability validation. The system reports aggregate costs (\$131.5 for 10 HTB machines; \$5.1 average per picoMini attempt) and discusses token conservation strategies with GPT-4-32k context limits.

Subsequent research addresses these limitations through complementary approaches: PenHeal~\cite{PenHeal2024} couples discovery with remediation using knapsack optimization but does not report token usage—the ``cost'' metric represents remediation scoring rather than LLM operational expenses. RefPentester~\cite{RefPentester2025} adds self-reflection and knowledge-guided planning, while browser-capable agents~\cite{BrowserPentestAgent2025} enable direct web interaction for CSRF/SSRF testing.

MAPTA advances autonomous security assessment through resource measurement and operational efficiency analysis that addresses fundamental gaps in prior work. Our evaluation provides complete token-level accounting across 104 XBOW challenges: 3.2M regular input, 1.10M output, 50.5M cached, and 0.595M reasoning tokens, totaling \$21.38 overall cost with median \$0.117 per challenge. This granular breakdown reveals output tokens as the primary cost driver, enabling resource optimization strategies unavailable prior.

Beyond cost accounting, MAPTA quantifies negative correlations between resource utilization and success—tool calls (r=-0.661), dollar cost (r=-0.606), tokens (r=-0.587), and time (r=-0.557)—providing actionable early-stopping heuristics and budget allocation guidance for production deployments. Our multi-agent architecture employs a coordinator/sandbox design with dynamic tool use, combined with end-to-end proof-of-concept validation that eliminates the false positives inherent in theoretical detection approaches. While prior systems discuss token pressure mitigation strategies, MAPTA measures and quantifies the complete operational profile, establishing the first rigorous cost-performance framework for autonomous penetration testing systems.  

\subsection{Benchmarks and Testbeds}

Traditional vulnerable applications (Juice Shop~\cite{JuiceShop}, WebGoat~\cite{WebGoat}, DVWA~\cite{DVWA}) focus on vulnerability types with implementations unsuitable for evaluating advanced systems. The XBOW benchmark dataset~\cite{XBOWBenchmark} represents significant advancement by providing modern web application challenges with REST APIs, complex business logic, and realistic authentication mechanisms. XBOW's key innovation emphasizes exploit execution validation over theoretical detection—each challenge requires actual exploitation success, eliminating false positives and aligning with real-world penetration testing objectives. Our approach builds on the fundamental insight from related work that effective automated security assessment requires tool orchestration, stateful reasoning, and practical verification~\cite{Yao2022ReAct,Atlidakis2019RESTler}. MAPTA's multi-agent architecture with sandboxed exploit validation directly addresses the limitations identified in single-agent systems like PentestGPT~\cite{Deng2024PentestGPT} and traditional scanners' false-positive challenges~\cite{Li2023SASTEvaluation}.

\section{Conclusion}

\textsc{MAPTA} demonstrates that multi-agent architectures can achieve competitive autonomous web application security assessment at practical scale. Our evaluation across 104 XBOW challenges achieves 76.9\% success with perfect performance on SSRF and misconfiguration vulnerabilities, while revealing systematic weaknesses in blind SQL injection (0\%) and cross-site scripting (57\%). The comprehensive cost accounting totaling \$21.38 establishes the first rigorous resource model for autonomous penetration testing, with median costs of \$0.073 for successful attempts versus \$0.357 for failures.

While our CTF evaluation (N=104) revealed strong correlations between resource usage and success (enabling early-stopping thresholds at approximately 40 tool calls, \$0.30, or 300 seconds), these patterns cannot be validated in our whitebox assessment due to the smaller sample size (N=10). Yet, MAPTA's real-world validation is impactful with 19 discovered vulnerabilities across ten popular open-source applications, of which \emph{14 classified as high or critical severity} (including RCE, command injections, secret exposure and arbitrary file write), at an average cost of \$3.67 per assessment. All findings are responsibly disclosed to the respective parties and bug bounty programs, where applicable. In total we are awaiting responses from 10 findings that are under CVE review. We expect that larger real-world scans will uncover substantially more vulnerabilities, and recommend deploying MAPTA on a continuous basis for immediate defensive action of web applications.


\clearpage

\appendix
\section*{Ethical Considerations}

The development and evaluation of MAPTA raises important ethical considerations regarding responsible disclosure of AI-powered security testing capabilities. We address these concerns through several key principles and safeguards implemented throughout our research.

\textbf{Defensive Publication and Community Awareness.} The primary ethical imperative for publishing this research stems from the reality that adversarial actors likely possess similar capabilities or are actively developing them. The democratization of AI development tools and the public availability of security testing methodologies means that malicious applications of these techniques are inevitable. By publishing our findings, we enable the cybersecurity community to understand and prepare for these emerging threats. Defensive security benefits from transparency about offensive capabilities, allowing organizations to implement appropriate countermeasures and security professionals to develop detection and mitigation strategies.

\textbf{Controlled Evaluation Environments.} Our evaluation methodology deliberately avoids testing against live production systems to prevent unintended harm or service disruption. We conducted two distinct types of assessments: (1) blackbox evaluation using purpose-built CTF challenges from the XBOW benchmark, which are explicitly designed for security testing and vulnerability discovery, and (2) whitebox assessments of open-source applications conducted entirely within isolated local environments. The whitebox evaluations involved cloning public repositories and conducting all testing within our own sandboxed virtual machines, ensuring no impact on production deployments or third-party infrastructure.

\textbf{Sandboxed Testing Infrastructure.} We implemented isolation measures to prevent any testing activities from affecting external systems. All MAPTA evaluations execute within dedicated virtual machines with restricted network access, preventing unintended outbound connections or data exfiltration. The sandbox environment includes monitoring and logging mechanisms to ensure all testing activities remain contained within the designated test boundaries. This approach eliminates risks of collateral damage while maintaining the authenticity of real-world vulnerability assessment scenarios.

\textbf{Responsible Vulnerability Disclosure.} For vulnerabilities discovered during our whitebox assessments, we follow responsible disclosure practices by notifying maintainers of affected projects through appropriate channels. The 10 vulnerabilities submitted for CVE assignment were reported to the respective project maintainers with sufficient detail for remediation while avoiding public disclosure of exploitation techniques until patches are available. We provide actionable remediation guidance and collaborate with maintainers to ensure timely resolution of identified security issues.

\textbf{Dual-Use Technology Considerations.} We acknowledge that MAPTA represents dual-use technology with both defensive and potentially offensive applications. To mitigate misuse risks, our implementation focuses on defensive security applications and includes built-in ethical constraints that prevent destructive operations, data exfiltration, or persistent system modifications. The system is designed to generate proof-of-concept demonstrations rather than weaponized exploits, providing sufficient evidence for vulnerability validation without enabling direct malicious use.

\textbf{Access Control and Distribution.} While we commit to making MAPTA source code publicly available upon publication to enable scientific reproducibility and defensive research, we implement responsible access controls. The release includes documentation emphasizing ethical use guidelines, configuration options for defensive-only operation modes, and integration with existing responsible security testing frameworks. We encourage adoption by legitimate security professionals, researchers, and organizations while discouraging malicious applications through community governance and ethical use agreements.

The fundamental ethical principle guiding this research is that the cybersecurity community benefits more from understanding these capabilities than from attempting to suppress them. As AI-powered development accelerates application creation, correspondingly advanced security assessment tools become essential for maintaining adequate security postures. MAPTA represents a defensive response to this challenge, providing organizations with capabilities to match the evolving threat landscape while adhering to responsible research and deployment practices.

\section*{Open Science \& Availability}
In accordance with the Open Science Policy, we provide complete access to all research artifacts necessary to evaluate and reproduce the contributions presented in this paper. All artifacts are available at \url{https://github.com/arthurgervais/mapta}. The updated XBOW 104 Challenge Evaluation Framework is available at \url{https://github.com/arthurgervais/validation-benchmarks}.

\bibliographystyle{plain}
\bibliography{main}

\end{document}